\documentclass[aps,prl,reprint,amsmath,superscriptaddress,showpacs]{revtex4-1}

\usepackage{graphicx}
\usepackage{dcolumn}
\usepackage{bm}

\begin{document}


\title{Collectivity evolution in the neutron-rich Pd isotopes towards the $N=82$ shell closure}

\author{H.~Wang}
 \email{wanghe@ribf.riken.jp}
\affiliation{School of Physics and State Key Laboratory of Nuclear Physics and Technology, Peking University, Beijing 100871, China}
\affiliation{RIKEN Nishina Center, 2-1 Hirosawa, Wako, Saitama 351-0198, Japan}

\author{N.~Aoi}
\affiliation{Research Center for Nuclear Physics, Osaka University, Ibaraki, Osaka 567-0047, Japan}

\author{S.~Takeuchi}
\affiliation{RIKEN Nishina Center, 2-1 Hirosawa, Wako, Saitama 351-0198, Japan}

\author{M.~Matsushita}
 \altaffiliation[Present address: ] {Center for Nuclear Study, University of Tokyo, RIKEN campus, Wako, Saitama 351-0198, Japan}
\affiliation{RIKEN Nishina Center, 2-1 Hirosawa, Wako, Saitama 351-0198, Japan}
\affiliation{Department of Physics, Rikkyo University, 3-34-1 Nishi-Ikebukuro, Toshima, Tokyo 172-8501, Japan}

\author{P.~Doornenbal}
\affiliation{RIKEN Nishina Center, 2-1 Hirosawa, Wako, Saitama 351-0198, Japan}

\author{T.~Motobayashi}
\affiliation{RIKEN Nishina Center, 2-1 Hirosawa, Wako, Saitama 351-0198, Japan}

\author{D.~Steppenbeck}
 \altaffiliation[Present address: ] {Center for Nuclear Study, University of Tokyo, RIKEN campus, Wako, Saitama 351-0198, Japan}
\affiliation{RIKEN Nishina Center, 2-1 Hirosawa, Wako, Saitama 351-0198, Japan}

\author{K.~Yoneda}
\affiliation{RIKEN Nishina Center, 2-1 Hirosawa, Wako, Saitama 351-0198, Japan}

\author{H.~Baba}
\affiliation{RIKEN Nishina Center, 2-1 Hirosawa, Wako, Saitama 351-0198, Japan}

\author{L.~C\'aceres}
\affiliation{Grand Acc\'el\'erateur National d'Ions Lourds, CEA/DSM-CNRS/IN2P3, Caen, France}

\author{Zs.~Dombr\'adi}
\affiliation{Institute for Nuclear Research, H-4001 Debrecen, Pf.51, Hungary}

\author{K.~Kobayashi}
\affiliation{Department of Physics, Rikkyo University, 3-34-1 Nishi-Ikebukuro, Toshima, Tokyo 172-8501, Japan}

\author{Y.~Kondo}
\affiliation{Department of Physics, Tokyo Institute of Technology, 2-12-1 Ookayama, Meguro, Tokyo 152-8551, Japan}

\author{J.~Lee}
\affiliation{RIKEN Nishina Center, 2-1 Hirosawa, Wako, Saitama 351-0198, Japan}

\author{K.~Li}
\affiliation{School of Physics and State Key Laboratory of Nuclear Physics and Technology, Peking University, Beijing 100871, China}
\affiliation{RIKEN Nishina Center, 2-1 Hirosawa, Wako, Saitama 351-0198, Japan}

\author{H.~Liu}
\affiliation{School of Physics and State Key Laboratory of Nuclear Physics and Technology, Peking University, Beijing 100871, China}
\affiliation{RIKEN Nishina Center, 2-1 Hirosawa, Wako, Saitama 351-0198, Japan}

\author{R.~Minakata}
\affiliation{Department of Physics, Tokyo Institute of Technology, 2-12-1 Ookayama, Meguro, Tokyo 152-8551, Japan}

\author{D.~Nishimura}
 \altaffiliation[Present address: ]{Department of Physics, Tokyo University of Science, 2641 Yamazaki, Noda, Chiba 278-8510, Japan}
\affiliation{Department of Physics, Osaka University, Toyonaka, Osaka 560-0043, Japan}

\author{H.~Otsu}
\affiliation{RIKEN Nishina Center, 2-1 Hirosawa, Wako, Saitama 351-0198, Japan}

\author{S.~Sakaguchi}
 \altaffiliation[Present address: ]{Department of Physics, Kyushu University, Fukuoka 812-8581, Japan}
\affiliation{RIKEN Nishina Center, 2-1 Hirosawa, Wako, Saitama 351-0198, Japan}

\author{H.~Sakurai}
\affiliation{RIKEN Nishina Center, 2-1 Hirosawa, Wako, Saitama 351-0198, Japan}
\affiliation{Department of Physics, University of Tokyo, 7-3-1 Hongo, Bunkyo, Tokyo 113-0033,Japan}

\author{H.~Scheit}
\affiliation{Insitiut f\"ur Kernphysik, Technische Universit\"at Darmstadt, 64289 Darmstadt, Germany}

\author{D.~Sohler}
\affiliation{Institute for Nuclear Research, H-4001 Debrecen, Pf.51, Hungary}

\author{Y.~Sun}
\affiliation{School of Physics and State Key Laboratory of Nuclear Physics and Technology, Peking University, Beijing 100871, China}

\author{Z.~Tian}
\affiliation{School of Physics and State Key Laboratory of Nuclear Physics and Technology, Peking University, Beijing 100871, China}

\author{R.~Tanaka}
\affiliation{Department of Physics, Tokyo Institute of Technology, 2-12-1 Ookayama, Meguro, Tokyo 152-8551, Japan}

\author{Y.~Togano}
 \altaffiliation[Present address: ] {Department of Physics, Tokyo Institute of Technology, 2-12-1 Ookayama, Meguro, Tokyo 152-8551, Japan}
\affiliation{RIKEN Nishina Center, 2-1 Hirosawa, Wako, Saitama 351-0198, Japan}

\author{Zs.~Vajta}
\affiliation{Institute for Nuclear Research, H-4001 Debrecen, Pf.51, Hungary}

\author{Z.~Yang}
\affiliation{School of Physics and State Key Laboratory of Nuclear Physics and Technology, Peking University, Beijing 100871, China}

\author{T.~Yamamoto}
\affiliation{Research Center for Nuclear Physics, Osaka University, Ibaraki, Osaka 567-0047, Japan}

\author{Y.~Ye}
\affiliation{School of Physics and State Key Laboratory of Nuclear Physics and Technology, Peking University, Beijing 100871, China}

\author{R.~Yokoyama}
\affiliation{Center for Nuclear Study, University of Tokyo, RIKEN campus, Wako, Saitama 351-0198, Japan}

\date{\today}

\begin{abstract}
The neutron-rich, even-even $^{122,124,126}$Pd isotopes has been studied via in-beam $\gamma$-ray spectroscopy at the RIKEN Radioactive Isotope Beam Factory. 
Excited states at 499(9), 590(11), and 686(17) keV were found in the three isotopes, which we assign to the respective $2^+_1 \rightarrow0^+_{gs}$ decays.
In addition, a candidate for the $4^+_1$ state  at 1164(20) keV was observed in $^{122}$Pd.
The resulting $E_x(2^+_1)$ systematics are essentially similar to those of the Xe ($Z=54$) isotopic chain and theoretical prediction by IBM-2,
suggesting no serious shell quenching in the Pd isotopes in the vicinity of $N=82$.
\end{abstract}



\maketitle

The existence of a shell structure is one of the most fundamental features of atomic nucleus.
Nuclei possessing the magic numbers, corresponding to the complete filling of shells, are relatively stable. 
Originally, the same set of magic numbers was thought to span over the entire nuclear chart
and a variation of protons and neutrons along isotonic and isotopic chains would not effect the magic nature of these nuclei.
Studies over the last two decades have, however, revealed that some numbers lose their
magicity in certain neutron-rich regions, while new magic numbers could arise.
For example, the nuclei around $^{32}$Mg are found 
by mass anomaly~\cite{Thibault_N=20}, the low first $2^+$ energy~\cite{Detraz_32Mg},
and large $\gamma$ decay probability~\cite{Moto_32mg} in $^{32}$Mg to exhibit a large collectivity despite the fact they are located near the $N=20$ shell closure.
Similar losses of magicity were found around $N=8$~\cite{Navin_12Be,Iwasaki_12Be_2+,Iwasaki_12Be_1-}
and $N=28$~\cite{Campbell_40Si_1, Bastin_42Si, Takeuchi_42Si}.

Recently the robustness of the $N=82$ shell closure attracts much attention, since it is also related to the nucleosynthesis in the
rapid neutron capture process ($r$-process)~\cite{Pfeiffer_rprocess}.
While $^{132}$Sn ($Z=50$, $N=82$) is known to exhibit typical characteristics of a doubly magic nucleus,  
such as a high-lying first excited state~\cite{132Sn_2+} and a sudden drop in neutron separation energy for the neighboring isotope~\cite{AME2012}, 
the structures of more proton-deficient nuclei are poorly known. 
Some spectroscopic information in the very neutron-rich region has been extracted
only for the Pd ($Z=46$) and Cd ($Z=48$) isotopes~\cite{Dillmann_130Cd_Qbeta,Jungclaus_130Cd,Walter_120Pd, Stoyer_120Pd}.
On the other hand, the nature of the $N=82$ magicity is intensively discussed based on various theoretical approaches~\cite{Grawe_N82, dobac_N82}.
Shell-model calculations including the tensor interaction suggest that a large $N=82$ shell gap is preserved~\cite{Grawe_N82}. 
In contrast, Hartree-Fock-Bogoliubov (HFB) calculations with the Skyrme force point towards a reduced gap 
due to a diffused potential caused by the large excess of neutrons~\cite{dobac_N82}. 
For the $r$-process, some mass models, such as those using HFB with Skyrme force SkP and ETFSI-$Q$ calculations,
assuming an $N=82$ shell quenching, are in better agreement with the abundance distribution around $A\approx130$~\cite{Chen_rprocess, Pearson_rprocess}.

The most neutron-rich nucleus of the $N=82$ isotones studied experimentally so far is $^{130}$Cd~\cite{Dillmann_130Cd_Qbeta,Jungclaus_130Cd}.
However, the conclusions of the previous reports contradict each other.
The $Q_{\beta}$ value was better reproduced by mass models assuming shell gap quenching~\cite{Dillmann_130Cd_Qbeta}.
On the other hand, the first $2^+$ state ($2^+_1$) was found at an excitation energy of 1.325 MeV, 
which is comparable to even-even $N=82$ isotones with $Z$ larger than 50~\cite{NNDC}
and therefore does not lead to a drastic modification to the shell closure.
The present study is aimed at adding more information on the nuclear structure in the vicinity of $N=82$ by extending the measurement of 
the energy of the first $2^+$ state ($E_x(2^+_1)$), to more proton deficient isotopes of palladium ($Z=46$), namely, $^{122,124,126}$Pd.

The experiment was performed at the RIKEN Radioactive Isotope Beam Factory (RIBF)
operated by the RIKEN Nishina Center and the Center for Nuclear Study of the University of Tokyo. 
To produce secondary cocktail beams via induced in-flight fission,
a $^{238}$U primary beam impinged on a 0.5 mm thick natural tungsten target located at the object point (F0) of the BigRIPS separator~\cite{Kubo_BigRIPS_1, Kubo_BigRIPS_2}.
The average primary beam intensity was 1.8~particle nA.
The momentum acceptance of BigRIPS was set to be 5\%. 
A cocktail beam composed of nuclei around $^{133}$Sn was analyzed and purified by the first stage of BigRIPS between the F0 and F3 foci 
using the magnetic rigidity ($B\rho$) selection in combination with an achromatic wedge-shaped energy degrader of 0.80 g/cm$^2$ thick aluminum located at the F1 dispersive focus.
The secondary beams were further purified by the second stage of BigRIPS between the F3 and F7 foci. 
A second $B\rho$ filter was employed by inserting a 0.40 g/cm$^2$ wedge-shaped aluminum degrader at the F5 dispersive focus. 
Ions passing through BigRIPS were identified event-by-event by measuring their $B\rho$, energy loss ($\Delta E$), and time-of-flight (TOF). 
The $B\rho$ value was obtained by trajectory reconstruction using the positions and angles measured at F3 and F5 by Parallel Plate Avalanche Counters (PPACs)~\cite{Kumagai_PPAC}.
The $\Delta E$ value was measured by an ionization chamber located at the F7 achromatic focus. 
The TOF was obtained from the time signal difference between two plastic scintillators located at F3 and F7, respectively. 
The atomic number $Z$ and mass-to-charge ratio $A/Q$ were obtained from $\Delta E$-TOF and $B\rho$-TOF correlations, respectively. 
Resolutions of 0.47 (FWHM) in $Z$ and 3.3$\times$10$^{-3}$ (FWHM) in $A/Q$ were obtained.
They were sufficient to clearly separate different isotopes. 
The total intensity of the secondary cocktail beams was 3$\times$10$^{4}$ particles per second,  $^{132,133}$Sn being the main composition.

To induce secondary reactions, the cocktail beams were incident on a 1.11 g/cm$^2$ thick $^9$Be secondary target placed at the F8 focus.
The average energy of the Sn isotopes at the center of the reaction target was around 230~MeV/nucleon. 
Reaction residues were delivered to the ZeroDegree spectrometer~\cite{Kubo_BigRIPS_2} for particle identification.
B$\rho$ values were set to maximize the transportation of $^{125}$Pd with a full momentum acceptance (8\%).
Various secondary reaction products were transported through ZeroDegree.
For Pd isotopes, the mass number of the transported ions ranges from $A=119$ to $A=127$.
After penetrating the reaction target, 90\% of the residues were in the fully stripped charge state.

The ZeroDegree particle identification was achieved in a similar way to the one for the secondary beam, by measuring $B\rho$, $\Delta E$, and TOF event-by-event.
To identify the charge states of the reaction residues, the total kinetic energy ($E$) was measured in addition. 
PPACs were located at the F8 and F11 achromatic and the F9 and F10 dispersive foci, respectively, while a second ionization chamber was placed at F11.
TOF was obtained from F8 to F11 by two plastic scintillators.
The reaction residue was stopped at F11 by a LaBr$_{3}$(Ce) scintillator to provide the $E$ value.
Particle identification was obtained from $\Delta E$-TOF, $B\rho$-TOF, and $E$-TOF correlations, respectively.
Resolutions in $Z$, $A/Q$, and $A$ were 0.47 (FWHM), 5.6$\times$10$^{-3}$ (FWHM), and 1.63 (FWHM), respectively.

De-excitation $\gamma$-rays emitted from the secondary reactions were measured by the DALI2 spectrometer~\cite{Takeuchi_DALI}.
It consisted of 186 large-volume NaI~(Tl) scintillation detectors surrounding the reaction target at polar angles from 14 to 148 degrees with respect to the beam direction. 
$\gamma$-ray energies measured in the laboratory system were Doppler-shift corrected based on the individual DALI2 detector angles.
The target chamber was covered by a 1~mm thick Pb shield to absorb low energy photons originating from atomic processes (mainly bremsstrahlung).
Energy resolution and efficiencies were estimated from Monte Carlo simulations using the GEANT4 framework~\cite{GEANT4}.
For a 1~MeV $\gamma$-ray emitted at a velocity of $v\approx0.6c$, they were 10\% (FWHM) and 20\%, respectively.

\begin{figure}
\includegraphics[width=8.6cm]{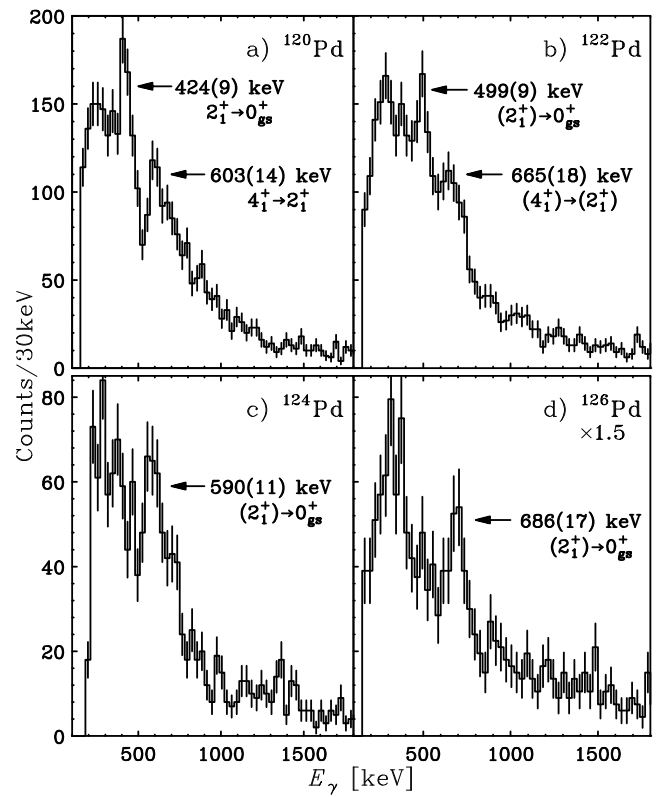}
\caption{\label{fig:pd_spectra}
DALI2 $\gamma$-ray energy spectra after correction for the Doppler-shift. 
The spectra in the different panels were obtained by applying coincidence gates  
with  a) $^{120}$Pd, b) $^{122}$Pd, c) $^{124}$Pd, and d) $^{126}$Pd
reaction residues detected in the ZeroDegree spectrometer.
}
\end{figure}

The Doppler-shift corrected energy spectra $\gamma$-rays emitted promptly with $^{120,122,124,126}$Pd isotopes are shown in Fig.~\ref{fig:pd_spectra}. 
In order to enhance the peak-to-background ratio, the $\gamma$-ray multiplicity detected in DALI2 was restricted to be lower than five for $^{120}$Pd to $^{124}$Pd,
but no multiplicity restriction was applied to the $^{126}$Pd data due to their lower statistics. 
For $^{120}$Pd, two transitions were observed at 424(9)~keV and 603(14)~keV, respectively,
which correspond to the known $2^+_1$ $\rightarrow$ $0^+_{\mathrm{gs}}$ and $4^+_1 \rightarrow 2^+_1$ decays at 438~keV and 618~keV~\cite{Walter_120Pd}.
The higher intensity for the $2^+_1$ $\rightarrow$ $0^+_{\mathrm{gs}}$ transitions is the expected behavior, 
as it collects feeding from higher-lying states populated by the fragmentation process.
The high-energy tail of the 603 keV peak presumably includes contributions from the $6^+_1 \rightarrow 4^+_1$ and $8^+_1 \rightarrow 6^+_1$ transitions at 738 and 795 keV 
reported in Ref.~\cite{Stoyer_120Pd}. 
The indicated errors for the peak energies include statistical and systematic uncertainties.
The latter originates from ambiguities in the energy calibration (2.5~keV), the shape of the background (3~keV), and $\gamma$-ray emission position 
as well as the velocity of the emitter caused by the unknown halflives of the excited states. 
The lifetime effect results in ambiguity in $\gamma$-ray emission angles and velocities used in the Doppler-shift correction (about 1\% of the $\gamma$-ray energy)~\cite{Pieter_lifetime}. 

For $^{122}$Pd, two new $\gamma$-ray transitions were observed at 499(9)~keV and at 665(18)~keV, respectively.
As the lower energy transition has a higher intensity and it is close in excitation energy to the known $2^+_1$ state in $^{120}$Pd, 
we assign the 499~keV peak to the  $2^+_1 \rightarrow 0^+_{\mathrm{gs}}$ transition.
By the same reason, we assign the second transition in $^{122}$Pd to the $4^+_1 \rightarrow 2^+_1$ decay.
Thus, a $E_x(4^+_1)$/$E_x(2^+_1)$ ratio ($R_{4/2}$) of 2.33(8) is obtained for $^{122}$Pd.

In the $\gamma$-ray spectra of $^{124}$Pd and $^{126}$Pd, peaks were observed at 590(11)~keV and 686(17)~keV, respectively,
which we assign to decays from the $2^+_1$ to the $0^+_{\mathrm{gs}}$ states.
In $^{124}$Pd, some enhancement is seen at around 710~keV, which could be contribution from the second excited state. 
This structure was taken into account in extracting the energy and its error for the $2^+_1$ state. 

\begin{figure}[th]
\begin{center}
  \includegraphics[width=8.6cm]{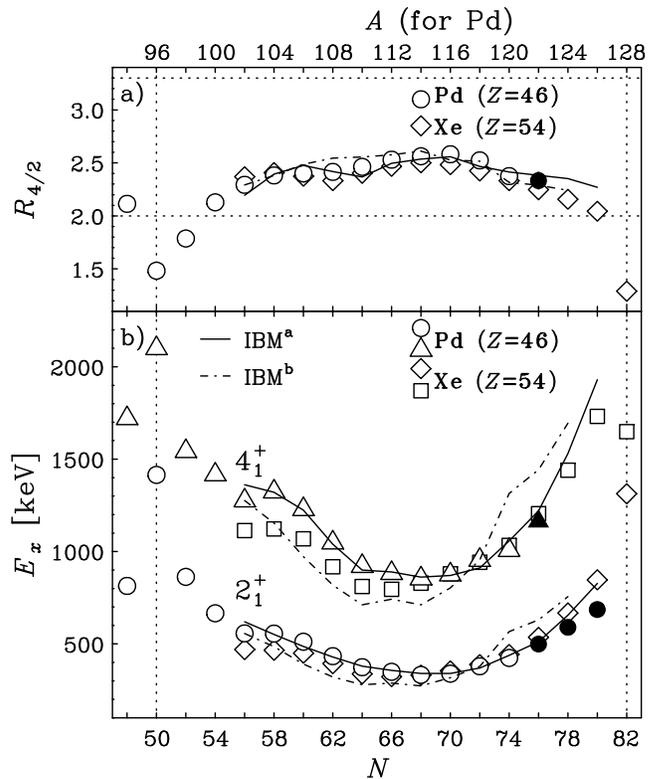}
\caption{\label{fig:pd_systematics}
In the upper panel 
a)  the $E_x(4^+_1)$/$E_x(2^+_1)$ ratios ($R_{4/2}$) are displayed as 
    a function of the neutron number for Pd (circles) and Xe (diamonds) isotopes.
    The horizontal dotted lines for values of 2.0 and 3.3 indicate 
    the vibrational and rotational limits, respectively.
The lower panel 
b)  shows $E_x(2^+_1)$ for Pd and Xe, and 
    $E_x(4^+_1)$ for Pd (triangles) and Xe (squares) isotopes as 
    a function of the neutron number. Open symbols and crosses were taken from Ref.~\cite{NNDC} and 
     filled symbols were obtained in this work. 
 Experimental error bars are smaller than the symbol sizes.
    Solid and dot-dashed lines display two sets of IBM-2 
    calculations~\cite{Kim_IBM2,Nomura_IBM2}.
   See text for details.   
 }
\end{center}
\end{figure}

Our new results extend, as shown in Fig.~\ref{fig:pd_systematics}, the systematics of the $2^+_1$ and $4^+_1$ excitation energies in the Pd isotopes to $N=80$ and $N=76$, respectively.
The $R_{4/2}$ ratios for the lighter mass even-even Pd isotopes~\cite{NNDC} as a function of the neutron number are also plotted in the figure.
While $^{96}$Pd features a large $2^+_1$ excitation energy due to the $N=50$ shell closure, isotopes in between the two major shells ($N=50$ and $N=82$) exhibit a parabolic trend
with a minimum at $N=68$.
Previous experiments already showed a gradual increase of the $2^+_1$ excitation energy commencing at $N=70$.
It is seen that the new data smoothly follow this trend up to $N=80$.

The $4^+_1$ excitation energies, as shown in Fig.~\ref{fig:pd_systematics} b), exhibit also a parabolic pattern 
similar to the one for $E_x(2^+_1)$, resulting in large $R_{4/2}$ ratios in the middle of the shell.
At $N=50$, the $R_{4/2}$ ratio is well below the vibrational limit (2.0), indicating the closed shell nature of $^{96}$Pd. 
The ratio increases as the neutron number and reaches about 2.5 in the middle of the shell, showing increased collectivity.
For Pd isotopes heavier than $A=116$, $R_{4/2}$ ratio decreases. 
The present result, 2.33(8) in $^{122}$Pd, follows this trend in the $R_{4/2}$ systematics.
These observations indicate that our new data follow the systematical trends pointing to diminishing collectivity towards $N=82$.

In Fig.~\ref{fig:pd_systematics}, data for the $Z=54$ (Xe) isotopes are plotted as well for comparison.
They are of four-proton particles instead of four-proton holes in Pd with respective to the $Z=50$ magic proton-core. 
In the $N_pN_n$ scheme~\cite{Casten_NpNn}, the four-proton-hole configuration and four-proton-particle configuration lead to 
the same signatures for nuclear structure, such as $E_x(2^+_1)$, $E_x(4^+_1)$, and $R_{4/2}$.
The systematics seen in Fig.~\ref{fig:pd_systematics} exhibit fair agreement between the Xe and Pd isotopes suggesting good particle-hole symmetry for protons and hence similar behaviors
in the neutron-number dependence for both the isotopes.
Considering the high $2^+_1$ energy at $^{136}$Xe ($N=82$), this similarity suggests that the Pd isotopes may not exhibit a strong shell quenching at $N=82$.

For the Pd isotopes, Kim $et$ $al$. and Nomura $et$ $al$. predict the $E_x(2^+_1)$ and $E_x(4^+_1)$ based on the Interaction Boson Model-2 (IBM-2)~\cite{Iachello_IBM2}.
They reported results by two approaches, one, noted IBM$^{a}$, with parameters determined from microscopic mapping calculations based on the known experimental energies of the even-even 
Pd isotopes from $^{102}$Pd to $^{114}$Pd~\cite{Kim_IBM2}, and the other, IBM$^{b}$, with a Hamiltonian derived from the Skyrme force mean-field~\cite{Nomura_IBM2}.
Both calculations, shown in Fig.~\ref{fig:pd_systematics}, well reproduce the overall experimental systematics.
Since the choice of the model space in these calculations corresponds to a good $N=82$ shell closure, the agreement supports the argument above, no strong shell quenching.

Slight hindrance of $E_x(2^+_1)$ in Pd isotopes from $N=76$ to $N=80$, as compared with the ones in Xe and the IBM-2 predictions, raises a question whether it points to a lower $E_x(2^+_1)$ at $N=82$ and hence certain weakening of the shell closure or to a high $E_x(2^+_1)$ value in $^{128}$Pd as in the Cd case.
Further studies are desired, for example, on $^{128}$Pd itself, which is identified recently at RIBF~\cite{Ohnishi_new2}.
Behaviors of neutron-rich nuclei with lower $Z$ are of interest, although their studies are experimental challenges with lower production cross sections.

In summary, low-lying excited states in $^{122,124,126}$Pd have been investigated via in-beam $\gamma$-ray spectroscopy at RIKEN RIBF with fast cocktail beams with mainly $^{132}$Sn and $^{133}$Sn.
It was found that the $E_x(2^+_1)$ value gradually increases in $^{122, 124, 126}$Pd  towards $N=82$. 
For $^{122}$Pd, a second excited state, which we assigned to $4^+_1$, was observed.
Comparisons with IBM-2 predictions and the data for Xe isotopes lead to the conclusion that a strong quenching of the $N=82$ shell gap is not likely to appear in Pd isotopes.

We express our gratitude to the accelerator staff of RIKEN Nishina Center for providing the intense $^{238}$U primary beam and to the BigRIPS team for tuning the secondary beam. 
The authors Z.D., D.So., and Z.V. were supported by OTKA Grants No K100835 and NN104543.

\bibliography{pd_submission}
\end{document}